\documentclass[12pt]{article}

\usepackage{amsmath}
\usepackage{amsthm}
\usepackage{graphics}
\usepackage{graphicx}
\usepackage{times}
\usepackage{bm}
\usepackage{natbib}
\usepackage{amsfonts}
\usepackage{enumerate, color}
\usepackage{setspace}
\usepackage{xspace}
\usepackage{lscape}
\usepackage{enumitem}
\usepackage{enumerate}
\usepackage{pdflscape}
\usepackage{afterpage}
\usepackage{multirow}

\usepackage{caption}
\usepackage{subcaption}

\usepackage{cleveref}

\usepackage{placeins} 

\usepackage[title]{appendix}

\usepackage{xr}
\usepackage{url} 

\newcommand{\blind}{1}

\def\A{\bm{A}} 
\def\B{\bm{B}}

\def\H{\bm{H}} 
\def\K{\bm{K}} 
 \def\m{\bm{m}} 
 
\def\P{\bm{P}}   
\def\R{\bm{R}} 
\def\S{\bm{S}} 

\def\V{\bm{V}} 
\def\U{\bm{U}}  
\def\W{\bm{W}} \def\w{\bm{w}} 
\def\X{\bm{X}} \def\x{\bm{x}} 
\def\Y{\bm{Y}}   
\def\Z{\bm{Z}}  

\def\bigO{\mathcal{O}}

\def\0{\bm{0}}  

\def\I{\bm{\mathcal{I}}} 

\def\cE{\mathrm{E}}

\def\var{{\rm{Var}}}

\def\det{{\rm{det}}}
\def \diag{{\rm{Diag}}}
\def\tr{{\rm{tr}}}
\def\ent{{\rm{ent}}}
\def\one{{\bm{1}}}
\def\zero{{\bm{0}}}

\def\balpha{\bm{\alpha}}

\def\bbeta{\bm{\beta}} 			 
\def\bdelta{\bm{\delta}} 		   
\def\btheta{\bm{\theta}}

\def\bOmega{\bm{\Omega}}	
\def\bSigma{\bm{\Sigma}}

\newtheorem{theorem}{Theorem} 
\newtheorem{corr}{Corollary} 
\newtheorem{assume}{\ignorespaces}

\begin{document}

\def\spacingset#1{\renewcommand{\baselinestretch}%
{#1}\small\normalsize} \spacingset{1}


\if1\blind
{
  \title{\bf Scalable Semiparametric Spatio-temporal Regression for Large Data Analysis}
  \author{
  	Ting Fung MA, 
    Fangfang WANG, 
    Jun ZHU, 
    \\ 
    Anthony R. IVES
    , and
    Katarzyna E. LEWI\'{N}SKA 
	}
  \maketitle
} \fi

\if0\blind
{
  \bigskip
  \bigskip
  \bigskip
  \begin{center}
    {\LARGE\bf Scalable Semiparametric Spatio-temporal Regression for Large Data Analysis}
\end{center}
  \medskip
} \fi

\bigskip
\begin{abstract}
With the rapid advances of data acquisition techniques, spatio-temporal data are becoming increasingly abundant in a diverse array of disciplines. Here we develop spatio-temporal regression methodology for analyzing large amounts of spatially referenced data collected over time, motivated by environmental studies utilizing remotely sensed satellite data. In particular, we specify a semiparametric autoregressive model without the usual Gaussian assumption and devise a computationally scalable procedure that enables the regression analysis of large datasets. We estimate the model parameters by quasi maximum likelihood and show that the computational complexity can be reduced from cubic to linear of the sample size. Asymptotic properties under suitable regularity conditions are further established that inform the computational procedure to be efficient and scalable. A simulation study is conducted to evaluate the finite-sample properties of the parameter estimation and statistical inference. We illustrate our methodology by a dataset with over 2.96 million observations of annual land surface temperature and the comparison with an existing state-of-the-art approach highlights the advantages of our method.

\end{abstract}


\noindent%


\vfill

\newpage
\spacingset{1.5} 

\section{Introduction}
\label{sec:intro}

With the rapid advances of data acquisition techniques, spatio-temporal data are becoming increasingly abundant in a diverse array of disciplines including the physical, biological, and social sciences \citep[see, e.g., ][]{cressie_wikle2011ST,dutilleul2011ST,anselin2013spatial}. Here we consider developing novel spatio-temporal regression methods for analyzing large amounts of spatially referenced data collected over time, motivated by environmental studies utilizing remotely sensed satellite data.

For illustration, we consider an environmental study of the land surface temperature (LST), which quantifies thermal energy flow among land surface, atmosphere and biosphere and thus, characterizes local ecological conditions. Changes in LST may have different causes, but the implications are critical for agriculture, biochemical processes, bioecology, economy, and health. 
The 2001-2019 MOD11A2 version 6 data, containing the night LST, were resampled to the 8km spatial resolution and the annual averages were computed for the contiguous US \citep{LSTdata}. 
It is of importance to investigate the time trend of LST across the study region while accounting for environmental conditions such as elevation and latitude. For example, the left column of Figure \ref{fig:Y_diff} in the Supplementary Materials shows that the LST has changed from 2001 to 2019, which arises from, among others, increase in global mean temperature \citep{NOAAlink} and change in land cover \citep{ref2_1,ref2_2}. In addition, it is of interest to examine LST in different ecoregions (Figure \ref{fig:ecoregion}) defined to be areas with similar landform, soil, vegetation, land use, wildlife, and hydrology \citep{Omernik2014}. 
Figure \ref{fig:ecoregion_top10}  in the Supplementary Materials shows the time series of LST in the ten largest Level III ecoregions, indicating that the LST time trend varies across different ecoregions.  

We could cast this research on LST in a spatio-temporal regression framework, regressing the response variable of LST on the predictor variables of time trend, ecoregion classes, and interactions between the time trend and ecoregions, as well as the environmental covariates of elevation and latitude. However, there are multiple challenges with using the existing spatio-temporal regression methods.
First, the sample size of the dataset is large.  With $T=19$ years and $N=155,900$ image pixels per year, there are well over 2.96 million LST observations in the dataset. The traditional spatio-temporal regression models with a regression mean and a spatio-temporal covariance function would be infeasible to implement, as the computations are on the order of $\bigO(N^3T^3)$ for evaluating the likelihood function and $\bigO(N^2T^2)$ for memory usage \citep[see, e.g., ][]{cressie93book,cressie_wikle2011ST}. 
There is ample room for innovations to reduce the computational burden and to make spatio-temporal regression analysis feasible for practical applications.

Second, although spatio-temporal statistics have advanced greatly in the past two decades, most of the state-of-the-art methods focus on the spatio-temporal dependence structure and the prediction of the underlying spatio-temporal processes. Raw data are often pre-processed by subtracting the mean function from the response (i.e., detrending) and the residuals are treated as the observed data. Even when the mean function is considered, the focus tends to be on the trend as a linear combination of the spatial coordinates and time points for the purpose of predicting the trend surface rather than the regression analysis \citep[see, e.g.,][]{cressie_etal_2019STR}. 
Thus these techniques are not directly applicable for the purpose of the LST study, calling for further research on the statistical inference of the mean function. 

Third, the distribution of the data is not necessarily Gaussian as is assumed by many existing spatio-temporal models. Indeed, the histograms depicted in Figure \ref{fig:hist_norm_curve} in the Supplementary Materials suggest a possible departure of the LST distribution from Gaussian.

There has been much research on the development of statistical methodology for analyzing spatio-temporal data \citep[see, e.g., ][]{huang_cressie1996,Zhang_etal2003,johannesson_etal2007,lu_etal2009,cressie_shi_kang2010,lee2015estimation,zhang_sang_huang2015,chu_zhu_wang2019}.
See also \cite{cressie_wikle2011ST} and \cite{cressie_etal_2019STR} for excellent reviews. 
For Gaussian processes, \cite{cressie_shi_kang2010}
proposed a fixed-rank filtering method for spatio-temporal data focusing on fast computation by dimension reduction spatially and fast smoothing, filtering, or forecasting over time, which in principle can be adapted to perform regression analysis but in practice is not quite feasible yet for the scale of our LST data.
\cite{Guinness2021} developed a Gaussian process (GpGp) method that scales up more readily and can be adapted to spatio-temporal regression analysis. GpGp type of methodology approximates the full likelihood of a Gaussian process by a product of conditional likelihoods on subsets, where the subsets are formed by reordering and grouping  the data \citep[see, e.g.,][]{Vecchia1988, guinness2018,statsciGp2021}. 
For non-Gaussian processes, \cite{chu_zhu_wang2019} and \cite{lee2015estimation} proposed semiparametric models without assuming Gaussian errors, which can be applied to spatio-temporal regression but both emphasized modeling the spatio-temporal dependence and the sample size needs to be kept at a modest size (in the thousands, not millions) for the methods to be computationally feasible. 
Alternatively, statistical modeling and inference are carried out under a Bayesian framework and the computational challenges are addressed by, for example, dimension reduction \citep{brynjarsdottir_berliner2014}, predictive processes \citep{banerjee2008} and Laplace approximation \citep{INLA, INLAreview}. 

Although the aforementioned statistical methods are useful for many applications, they remain to be either infeasible when the sample size is on the scale of our LST data and/or not well suited for regression analysis involving spatio-temporal processes that are not necessarily Gaussian. 
Thus here we aim to develop a novel computationally scalable procedure that enables the regression analysis of large datasets, while guided by rigorous asymptotic theory and computational complexity analysis.
For modeling and theoretical development, our proposed method is semiparametric in the sense that no explicit distributional assumption is made and only finite-moment conditions of the underlying spatio-temporal processes are assumed. 
We estimate the model parameters by maximizing a quasi-likelihood and establish the asymptotic properties of the parameter estimators. 
In addition, the existing literature in spatial statistics tends to directly model the spatio-temporal covariance function and provide its approximation \citep[see, e.g., ][]{cressie_huang1999,gneiting2002,chu_zhu_wang2019}. 
Here we take an alternative approach to modeling the spatio-temporal dependence by autoregression. While the autoregression modeling idea is widely used particularly in econometrics, most existing methods are not computationally scalable to the size of our LST data \citep[see, e.g., ][]{lee2015estimation,chi_zhu2019book}. Although \cite{Guo2016} and \cite{Gao2019} considered the estimation and inference of autoregressive models under the high-dimensional setting, the methodology focuses on the estimation of coefficient matrices for zero-mean autoregressive processes without addressing the regression nor computational complexity in detail. 
In terms of computation, our proposed procedure can be used for spatio-temporal regression with a large sample size $NT$ as in the LST data example. In particular, we adopt advanced computational techniques including efficient data pre-processing, constrained sequential quadratic programming (SQP), and implicit parallel computing. 
These computational innovations provide a substantive improvement over the existing methods in the literature that tend to require tuning parameter estimation and/or approximating the spatio-temporal neighborhood structure for statistical inference \citep[see, e.g.,][]{INLA,Bai2012,guinness2018,Guinness2021,Zhao2021}. 

The remainder of this paper is organized as follows. Section~\ref{sec:model} presents the model and its estimation. Section~\ref{sec:theory} establishes the asymptotic properties of the quasi maximum likelihood estimates of the model parameters. Section~\ref{sec:compute} provides a fast computational procedure for estimation and inference. The finite-sample properties of the estimators are assessed by simulation studies in Section~\ref{sec:sim} and the LST data example is given in Section~\ref{sec:realdata}. 
Proofs of the theoretical results and other technical details including additional computational aspects, tables, and figures are provided in the Supplementary Materials.

\section{Model and Estimation}
\label{sec:model}

\subsection{Model Specification}

At time $t\in\mathbb{Z}$, let $\Y_t= (Y_{1,t},\ldots,Y_{N,t})'$ denote an $N$-dimensional vector that contains the response variables from $N$ cells that partition the study region of interest in $\mathbb{R}^2$. Let $\X_{t}$ denote an $N\times k$ design matrix of $k$ nonstochastic predictor variables. We model the spatio-temporal evolution of $\Y_t$ in relation to the design matrix $\X_t$ through the following spatio-temporal regression model 
\begin{equation}\label{eqn.y1}
	\begin{array}{cccccccccc}
		\Y_t          &=&  \X_{t}\bbeta                    & + &     \U_t, & t \in\mathbb{Z}, 
	\end{array}
\end{equation}
where $\bbeta$ denotes a $k\times 1$ vector of regression coefficients. 
The spatio-temporal error $\U_t$ is stochastic and modeled by a spatio-temporal dynamic process such that  
\begin{equation}\label{eqn.u1}
	\begin{array}{ccccccccccc}
		\U_t          &=& \lambda \W \U_t  & + & \rho \W \U_{t-1} & + & \gamma \U_{t-1}  & + &    \V_t,      
	\end{array}
\end{equation}
where $ \V_t = (v_{1,t}, \ldots, v_{N,t})'$ is an $N\times 1$ vector of innovations that are assumed to be {\it iid}, not necessarily Gaussian, with mean zero and variance $\sigma^2 \I_{N}$ and $\I_{N}$ is the $N\times N$ identity matrix. The spatio-temporal dependence parameters include the conventional temporal lag effect $\gamma$, the contemporaneous spatial interactions effect $\lambda$, and the effect of spatial diffusion that takes place over time $\rho$ \citep[see, e.g., ][]{anselin2013spatial,lee2015estimation,chi_zhu2019book}. 

Finally, the spatial weight matrix $\W$ is an $N\times N$ nonstochastic symmetric matrix with zero diagonals for a given spatial neighborhood structure \citep{cressie93book}.  The symmetry of $\W$ has important implications on computation, which will be elaborated in later sections.  Special cases of the spatial weight matrix $\W$ include the block-diagonal structure and commonly assumed first- or second-order neighborhood structures. For a block-diagonal structure, $\W = \diag\{\w_1,\ldots,\w_p\}$, where $\w_i$ is a $n_i\times n_i$ matrix, with $N=\sum_{i=1}^pn_i$ \citep{case1991spatial}.  On a regular square grid, the first-order neighbors are the four nearest cells whereas the second-order neighbors are the eight nearest cells \citep{cressie93book}. 
In addition, the spatial weight matrix could be used to construct the design matrix $\X_t$ in order to capture the spatial neighboring effects; 
for instance, let $\X_t = (\one_{N}, \Z_{1t}, \Z_{2t})$, where $\Z_{2t} = \tilde\W\Z_{1t}$ and $\tilde\W$ is a spatial weight matrix defined above. 

Let  $\btheta = (\lambda, \gamma, \rho)'$ denote the vector of the spatio-temporal dependence parameters. We define $\R(\btheta) = \rho \W + \gamma\I_{N}$, $\S(\lambda) = \I_{N} -\lambda \W$, and
$\A(\btheta) =  \R(\btheta) \S(\lambda)^{-1}.$
We may then rewrite the spatio-temporal dynamic process \eqref{eqn.u1} as
$ 
\S(\lambda) \U_t  =  \A(\btheta) \S(\lambda) \U_{t-1}  +  \V_t.
$ 
That is, the spatio-temporal error $\U_t$ follows a vector autoregression model of order one and can be shown to be weakly stationary under the assumption that $\S(\lambda)$ is non-singular and the eigenvalues of $A(\btheta)$ are all strictly less than one in magnitude. 

\subsection{Parameter Estimation}

For the observed response vectors $\Y_1, \ldots, \Y_T$ modeled by \eqref{eqn.y1},  we define the vector of all the spatio-temporal errors $\U  = (\U_1', \U_2', \ldots, \U_T')'$ and its matrix operator
\begin{equation}\label{eqn.B} 
	\B (\btheta) = \left(
	\begin{array}{cccccccccc}
		\S(\lambda)  & 0 &  \cdots &    0 & 0 \\
		-\R(\btheta) & \S(\lambda) &    \cdots & 0 & 0 \\
		\vdots & \vdots &    \vdots & \vdots & \vdots \\
		0 & 0 &    \cdots & -\R(\btheta) & \S(\lambda)  
	\end{array}
	\right)_{NT\times NT}
\end{equation}
such that $\B(\btheta) \U  = (( \S(\lambda) \U_1)', \V_2', \ldots, \V_T')'$. The covariance matrix of $\B(\btheta) \U$ is $\sigma^2\bOmega(\btheta)$, where $\bOmega(\btheta)  = \diag(\K(\btheta), \I_{N}, \ldots, \I_{N})$ and  $\K(\btheta) = \sum_{j=0}^{\infty} \A(\btheta)^{j} \A(\btheta)^{'j}$.  

Let $\bdelta = (\bbeta', \btheta', \sigma^2)'$ denote the vector of all the model parameters.  To estimate $\bdelta$, we propose the following quasi log likelihood function,
\begin{align} 
	\log L_{NT}(\bdelta) =& - \frac{NT}{2}\log(2\pi\sigma^2)- \frac{1}{2} \log \det(\K(\btheta)) + T\log |\det(\S(\lambda))|   \nonumber\\
	& - \frac{1}{2\sigma^2} (\Y-\X\bbeta)'\bSigma(\btheta)^{-1}(\Y-\X\bbeta), 
	\label{eqn.llk}
\end{align}	
where $\Y = (\Y_1', \ldots, \Y_T')'$ denotes the $NT\times 1$ vector of all the response variables, $\X = (\X_1', \ldots, \X_T')'$ is the corresponding $NT\times k$ design matrix, and $\bSigma(\btheta)^{-1}  = \B (\btheta)'(\bOmega(\btheta))^{-1}\B (\btheta)$ is the precision matrix. 
Denote by $\widehat{\bdelta}$  the maximizer of the quasi log likelihood function $\log L_{NT}(\bdelta)$; that is,  
$$
\widehat{\bdelta} = \arg\max_{\bdelta\in \Theta_{\bdelta}} \log L_{NT}(\bdelta), 
$$
where $\Theta_{\bdelta}$ is the parameter space specified in Appendix \ref{sec:assume}.  Throughout this paper, we refer to $\widehat{\bdelta}$ as our quasi-maximum likelihood estimator (QMLE) of the model parameters $\bdelta$.  

As illustrated by the LST data example in Section \ref{sec:intro}, our primary interest is statistical inference of the regression coefficients $\bbeta$, while the estimation of the spatio-temporal dependence parameters $\btheta$ is of secondary interest intended to account for spatio-temporal correlation when drawing the inference about $\bbeta$.

\section{Asymptotics}\label{sec:theory}


Under suitable regularity conditions,  we may establish the asymptotic properties of the QMLE $\widehat\bdelta$, as the number of cells $N\rightarrow \infty$ while the number of time points $T$ can be either fixed or $T\rightarrow\infty$. 
Denote by $\bdelta_0=(\bbeta_0', \btheta_0', \sigma_0^2)'$ the vector of true model parameters. We first consider the case that $T$ is fixed.

\begin{theorem}\label{thm.consistency}
	Suppose that $\W$ has more than two distinct eigenvalues, and Assumptions \eqref{assume.W}--\eqref{assume.sigma} hold. Then,
	 $\bdelta_0$ is identifiably unique and $\widehat\bdelta\overset{p}{\longrightarrow}\bdelta_0$ as $N\rightarrow\infty$. 
\end{theorem}

Theorem~\ref{thm.consistency} establishes that the QMLE $\widehat{\bdelta}$ is a consistent estimator of the true parameter vector $\bdelta_0$ in the sense that $\widehat\bdelta$ converges to $\bdelta_0$ in probability, when $N\rightarrow \infty$. 

Next, under additional conditions about the higher-order properties of the quasi log likelihood function, we examine the asymptotic distribution of the QMLE $\widehat{\bdelta}$.

\begin{theorem}\label{thm.clt}
Suppose that the conditions in Theorem \ref{thm.consistency}, and additional Assumptions 
\eqref{assume.UB2} and \eqref{assume.hessian} are fulfilled.	Then,
\begin{equation}\label{eqn:clt.delta}
	\sqrt{N}(\widehat\bdelta-\bdelta_0) \overset{d}{\longrightarrow} \mathcal{N}\left(\zero, 4\left\{\overline\bSigma_{1}^{-1}  +  \overline\bSigma_{1}^{-1}\overline\bSigma_{2}\overline\bSigma_{1}^{-1}\right\} \right),
\end{equation}
where  $\overline\bSigma_{1} = \lim_{N\rightarrow\infty} N^{-1}\bSigma_{1, N}$, $\overline\bSigma_{2} = \lim_{N\rightarrow\infty} N^{-1}\bSigma_{2, N}$,  $\bSigma_{1, N}  = \diag(4 \sigma_0^{-2}\X'\bSigma(\btheta_0)^{-1}\X, 2\bOmega_{N} )$ with $\bSigma(\btheta_0)^{-1}  = \B (\btheta_0)'(\bOmega(\btheta_0))^{-1}\B (\btheta_0)$ and $\bOmega_{N} $ defined in \eqref{eqn.bSigma1N}, and $\bSigma_{2, N}$ is defined in \eqref{eqn.bSigma2N} in the Supplementary Materials. 
\end{theorem}

Theorem~\ref{thm.clt} establishes that $\widehat{\bdelta}$ converges to a multivariate Gaussian distribution at the rate of $\sqrt{N}$. The asymptotic covariance matrix involves two matrices $\overline\bSigma_1$ and $\overline\bSigma_2$, which can be replaced by their consistent estimators for the evaluation of the asymptotic distribution of $\widehat{\bdelta}$ in practice. Since the primary interest is in the statistical inference about the regression coefficients $\bbeta$, we present the asymptotic distribution of $\widehat{\bbeta}$ and its relationship to the other parameter estimators $\widehat{\btheta}$ and $\hat{\sigma}^2$ in the following corollary.

\begin{corr} \label{cor.beta}
	Suppose the conditions in Theorem \ref{thm.clt} hold. Then, we have
	\begin{equation} \label{eqn:clt.beta}
		\sqrt{N}(\widehat\bbeta-\bbeta_0) \overset{d}{\longrightarrow} \mathcal{N}\left(\zero,  \overline\bSigma_{\bbeta_0}^{-1}\right),	
	\end{equation}
	where $\overline\bSigma_{\bbeta_0} = \sigma^{-2}_0\lim_{N\rightarrow\infty} N^{-1}\X'\bSigma(\btheta_0)^{-1}\X$. Under the additional assumption that $\mu_3 = \cE (v_{j,t}^3)=0$, $\widehat{\bbeta}$ is asymptotically independent of $\widehat{\btheta}$ and $\hat{\sigma}^2$. 
	
\end{corr}

 
Corollary~\ref{cor.beta} provides the basis for a computationally efficient approach to the statistical inference about the regression coefficients $\bbeta$, as we will detail in Section \ref{subsec:estcomplexity}.
Moreover, the asymptotic distribution of $\widehat{\bbeta}$ remains unchanged regardless of the distribution of the spatio-temporal innovations.  In particular, when the innovation is symmetric (i.e., $\mu_3=0$, which is satisfied by many commonly used distributions including Gaussian and Student-$T$ distributions),  Corollary~\ref{cor.beta} establishes that $\widehat{\bbeta}$ is asymptotically independent of the spatio-temporal dependence parameter estimators $\widehat{\btheta}$ and the variance component estimator $\hat{\sigma}^2$.  

By Theorems \ref{thm.consistency} and \ref{thm.clt},  $N^{-1}\bSigma_{1, N}$ and  $N^{-1}\bSigma_{2, N}$ with $\bdelta_0$ replaced  $\widehat{\bdelta}$ converge in probability to $\overline\bSigma_{1}$ and  $\overline\bSigma_{2}$, respectively, as $N\rightarrow\infty$.  Thus,  a consistent estimator of the asymptotic covariance matrix of $\widehat\bdelta$  can be obtained from
\begin{equation}\label{eqn.asymcovdelta}
	4\left\{ \bSigma_{1, N}^{-1}  +  (\bSigma_{1, N})^{-1}(\bSigma_{2, N})(\bSigma_{1, N})^{-1}\right\} 
\end{equation}
evaluated at the QMLE $\widehat\bdelta$.  However, $\bSigma_{1, N}$ and $\bSigma_{2, N}$ are both challenging to compute when $N$ is large. One major challenge is that the calculation of $\bSigma_{1,N}$ and $\bSigma_{2,N}$ requires solving large linear systems, or equivalently inverting large matrices, which is computationally expensive. On the other hand, the estimator of the upper-left block of $\bSigma_{1, N}$ is readily available as $\sigma_0^{-2}\X'\bSigma(\btheta_0)^{-1}\X$ can be consistently estimated by $\hat{\sigma}^{-2}\X'\bSigma(\widehat{\btheta})^{-1}\X$. Thus, a consistent estimator of the asymptotic covariance matrix of $\widehat\bbeta$ is $\hat\sigma^{2}(\X'\bSigma^{-1}(\widehat\btheta)\X)^{-1}$, and its computation can in fact be made scalable (see Section \ref{sec:compute}).  

In addition, the asymptotic results hold when $T$ is either fixed or tends to infinity with $N$ at an arbitrary rate. That is, Theorems~\ref{thm.consistency}, \ref{thm.clt} and Corollary~\ref{cor.beta} can be readily extended to the case when $N$ and $T$ both tend to infinity, in which case the rate of convergence in Theorem~\ref{thm.clt} and Corollary~\ref{cor.beta} becomes $\sqrt{NT}$ instead of $\sqrt{N}$, with corresponding adjustment of Assumptions \eqref{assume.x}, \eqref{assume.sigma} and \eqref{assume.hessian}, and the asymptotic covariance matrices.

Before closing this section, we remark on the regularity conditions \eqref{assume.W}--\eqref{assume.hessian} provided in the Appendix. 
As assumed in \eqref{assume.W}, the spatial weight matrix is symmetric; for example, it is common to set the weight between two distinct locations to one if the distance between the two locations are within a certain threshold and zero otherwise \citep[see, e.g.,][]{cressie93book}. The symmetry assumption plays a vital role in facilitating the computation, because it follows that $\S(\lambda)$ and $\R(\btheta)$ are symmetric as well. Further, the long-run covariance matrix of the process $\S(\lambda) \U_t $, $\K(\btheta)$, can be written as $(\I_N - \A(\btheta)^{2})^{-1}$, thereby reducing $\S(\lambda)\K(\btheta)^{-1}\S(\lambda)$ to $\S(\lambda)^2-\R(\btheta)^2$, which enables efficient computation of the quasi log likelihood function \eqref{eqn.llk} (see Section \ref{subsec:estcomplexity} for details). In contrast, without the symmetry assumption, the long-run covariance matrix involves infinite terms approximated by a sum of finite matrices by assuming that the process starts at some time point in the past instead of the infinite past \citep[see, e.g.,][]{lee2015estimation}, which not only poses a challenge for evaluating the quasi log likelihood function in practice when the sample size is large, but also makes the resulting error process possibly non-stationary.


For Assumption \eqref{assume.S}, a sufficient condition for the matrix $\S(\lambda)=\I_N -\lambda \W$ being non-singular and the eigenvalues of $A(\btheta)$ being less than one in magnitude is that the parameters $\lambda, \gamma, \rho$ satisfy the following inequality: 
\begin{equation} \label{eqn.parameter_space}
	(\lambda^2 - \rho^2)d_{j}^2 - 2(\lambda+\gamma\rho)d_{j} + (1-\gamma^2) >0, j=1,\ldots,r ,
\end{equation}
where $\{d_i, i=1,\ldots,r\}$ are the non-zero eigenvalues of $\W$ with the smallest eigenvalue ($d_1$) and the largest eigenvalue ($d_r$) of $\W$ having opposite signs. For \eqref{eqn.parameter_space} to hold, it is sufficient to consider the following set,
\begin{equation}\label{eqn.suff.para.space}
	\left\{(\lambda, \gamma, \rho): -1 < \gamma< 1, \frac{1-\gamma}{d_1} < \lambda + \rho < \frac{1-\gamma}{d_r}, \frac{1+\gamma}{d_1} < \lambda - \rho < \frac{1+\gamma}{d_r}\right\}.
\end{equation}
In practice, we choose $\mathbf{\Theta}_{\btheta}$ as a compact subset of the above set. 

Assumption \eqref{assume.V} assumes that the innovations are independent and identically distributed over time and across space and requires the existence of unconditional higher-order moments without assuming a specific distribution. Assumption \eqref{assume.UB} ensures that  $N^{-1}(\ell_{N}(\bdelta) - \cE\ell_{N}(\bdelta)) = o_p(1)$ where $\ell_{N}(\bdelta)=-2 \log L_{NT}(\bdelta)$, 
which is weaker than the condition in the literature as we do not assume $\S(\lambda)$ to be positive definite \citep[see, e.g.,][]{lee2015estimation}.  Assumption \eqref{assume.x} ensures the non-singularity of $\overline\bSigma_{1}$ when $N\rightarrow\infty$. By assuming that  $\W$ has more than two distinct eigenvalues along with Assumptions \eqref{assume.x} and \eqref{assume.sigma}, $\liminf_{N\rightarrow\infty}N^{-1} \{\cE(\ell_{N}(\bdelta)) - \cE(\ell_{N}(\bdelta_0))\}$ $>0$  for $\bdelta \neq \bdelta_0$, which implies $\bdelta_0$ can be uniquely identified \citep[see, e.g.,][]{gallant1988unified}.  Assumption \eqref{assume.UB2} regulates the gradient and the Hessian matrix of $\ell_{N}(\bdelta)$ around $\bdelta_0$, while Assumption \eqref{assume.hessian} ensures that the asymptotic covariance matrix of $\widehat{\bdelta}$ is well-defined when $N\rightarrow\infty$, both of which are standard regularity conditions. 

%
%
\section{Computation} \label{sec:compute}

In this section, we will develop a novel fast computation procedure and show that its computational complexity is on the order of $\bigO(NT)$ for obtaining the QMLE $\widehat{\bdelta}$ and the variance estimate of $\widehat{\bbeta}$, which is scalable to the sample size of the LST data. The existing state-of-the-art methodology generally approximates the dependence structure for computational ease, while our approach does not require an approximation of the spatio-temporal dependence. Thus, our computational procedure provides a novel and scalable alternative to the existing spatio-temporal modeling and inference without approximating the likelihood function.
%
%
%

\subsection{Computational Procedure}\label{sec.compalgo}

\begin{figure}[hp]
	\centering
	\includegraphics[width=1\textwidth]{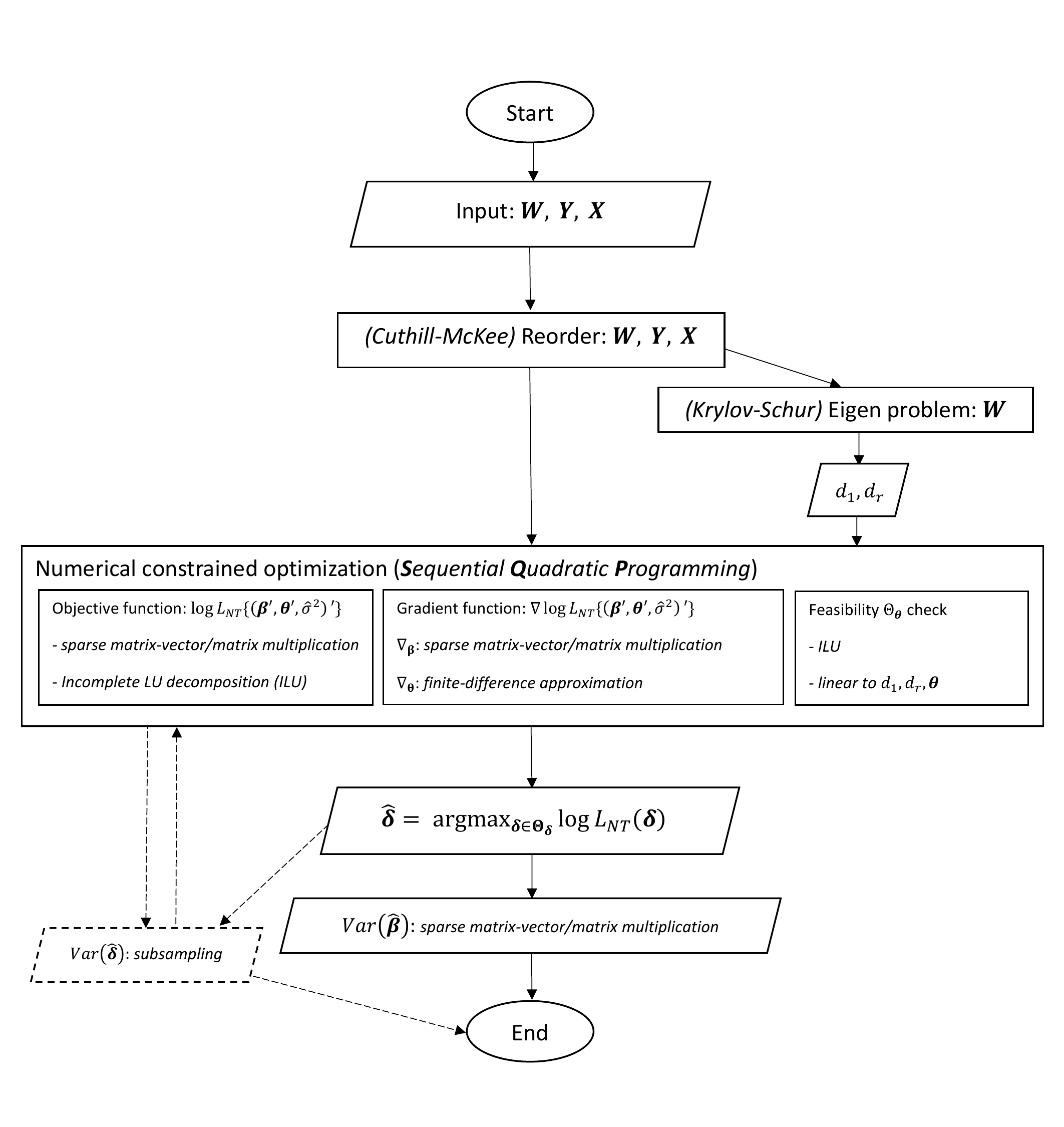}
	\caption{Flowchart for carrying out the proposed spatio-temporal regression and inference.}
	\label{fig:flowchart}
\end{figure}

We obtain the QMLE, $\widehat{\bdelta}$, and an estimate of its variance $\var(\widehat{\bdelta})$ by bringing together a set of computational techniques for nonlinear optimization and sparse matrix operations.  An overview of the procedure is visualized by a flowchart in Figure \ref{fig:flowchart}. Specifically, the procedure starts with the input of the spatial weight matrix $\W$, the vector of the response variables $\Y$, and the design matrix $\X$. We then preprocess the data by applying the reverse Cuthill-McKee (RCM) algorithm \citep{Gilbert92sparsematrices}. In particular, the RCM algorithm permutes the rows and columns of $\W$, which is a symmetric, generally sparse matrix, into a symmetric sparse banded matrix with a small bandwidth. This effectively moves the non-zero elements of $\W$ towards the diagonal while preserving the spatial neighborhood structure. The underlying graph theory for the RCM algorithm views the spatial weight matrix as a graph with vertices (of spatial locations) and edges that connect spatial neighbors specified in the $\W$ matrix. We then reorder the rows of $\Y_t$ and $\X_t$ according to the Cuthill-McKee ordering of $\W$ for $t=1,\ldots,T$. For a given spatial weight matrix $\W$, it is always possible to obtain a sparse banded $\W$, without distorting the pre-specified spatial neighboring structure \citep{bandedsurvey}. Thus henceforth we assume that $\W$ is a pre-specified symmetric sparse banded matrix with bandwidth $b$, which eases the implementation of computational techniques for banded matrices and enables a more precise account of computational complexity. 

Next, the parameter vector $\bdelta$ is estimated by maximizing the quasi log likelihood using an iterative SQP (i.e., fmincon() in MATLAB) \citep[see Chapter 18 of][]{NocedalWright2006}. At each iteration, the quasi log likelihood function \eqref{eqn.llk} and its gradient functions are evaluated for optimizing \eqref{eqn.llk} subject to a set of constraints on the parameter space $\bm{\Theta}_{\bdelta}$. To ensure the scalability of SQP, however, care is needed in the evaluation of the quasi log likelihood function, as we will show in the next subsection. 
In addition, the constraints on the parameters need to be checked, which we will refer to as feasibility check.  The standard feasibility check would require computational cost on the order of $\bigO(N^{2.4})$. 
Here we apply the sufficient condition \eqref{eqn.suff.para.space} developed in Section~\ref{sec:theory}, which requires solving for the smallest ($d_1$) and largest ($d_r$) eigenvalues of $\W$. We thus preprocess $\W$ by the Krylov-Schur algorithm, which is an iterative method for solving eigenproblems with sparsity and belongs to the class of Krylov subspace methods \citep{Stewart2002}. The Krylov-Schur algorithm first generates a sequence of subspaces containing the approximations of a subset of eigenvectors and eigenvalues of $\W$. Then these approximations are extracted by applying the QR algorithm to the projection of $\W$ onto the subspaces and the subset of eigenvalues is approximated iteratively through the Arnoldi method. A reordering of the Schur decomposition in the previous step is considered to improve the standard Arnoldi method \citep[see Chapter 3 of][]{eigenbook2005}. Based on $d_1$, $d_r$, and the sufficient condition \eqref{eqn.suff.para.space}, our feasibility check requires $\bigO(1)$ operations, which is a significant improvement over the $\bigO(N^{2.4})$ operations and the $\bigO(N)$ memory usage when a full eigen-decomposition of $\W$ is used for \eqref{eqn.parameter_space}.


In addition, most of the computations can be parallelized and in particular, we enable the implicit parallelism through maxNumCompThreads(), which distributes the computation in multiple cores and utilizes the sparsity of matrices in our MATLAB code \citep{MATLABparallel}. Similar techniques can also be implemented in R and Python, for example through the Basic Linear Algebra Subroutines (BLAS) or Linear Algebra Package (LAPACK) \citep{lapack99,blackford2002updated,BulucGilbert2011}.

\subsection{Computational Complexity} \label{subsec:estcomplexity}

Direct evaluation of the quasi log likelihood function \eqref{eqn.llk} requires $\bigO(N^3T^3)$  
operations and is computationally infeasible when $NT$ is large. In the following we show that our computational procedure has the computational complexity of $\bigO(NT)$ for the regression coefficient estimation and inference, which is linear to the sample size and thus is scalable to the size of the LST dataset. 

Note that the matrix operator $\B(\btheta)$ in \eqref{eqn.B} and the covariance matrix $\bOmega(\btheta)$ are involved in the quasi log likelihood function \eqref{eqn.llk}. Storing the entirety of $\B(\btheta)$ and $\bOmega(\btheta)$ during the process of computing the precision matrix would require standard memory usage and operations to be on the order of $\bigO(N^2T)$. Instead, we partition the matrix operator $\B(\btheta)$ and the covariance matrix $\bOmega(\btheta)$ in such a way that we only store the unique non-zero blocks of quadratic terms. More specially, we rewrite the last term of the quasi log likelihood function \eqref{eqn.llk} as
\begin{eqnarray} 
	&& (\Y-\X\bbeta)'\bSigma(\btheta)^{-1}(\Y-\X\bbeta)= (\Y-\X\bbeta)'\B (\btheta)'(\bOmega(\btheta))^{-1}\B (\btheta)(\Y-\X\bbeta)\label{eqn.llksum} \\
	&=&  (\Y_1  - \X_{1}\bbeta)'\S(\lambda)\K(\btheta)^{-1}\S(\lambda) (\Y_1  - \X_{1}\bbeta) + \sum_{t=2}^{T}(\Y_t  - \X_{t}\bbeta)'\S(\lambda)^2(\Y_t  - \X_{t}\bbeta) \nonumber\\
	&& +  \sum_{t=1}^{T-1}(\Y_t  - \X_{t}\bbeta)'\R(\btheta)^2(\Y_t  - \X_{t}\bbeta)   -2\sum_{t=1}^{T-1}(\Y_t  - \X_{t}\bbeta)'\R(\btheta)\S(\lambda) (\Y_{t+1} - \X_{t+1}\bbeta).\nonumber
\end{eqnarray}
Note that the total number of non-zero elements ($nnz$) of $\W$ is $\bigO(bN)$.
Since the product of two $N\times N$ banded matrices each with bandwidth $\bigO(b)$ is still  banded  with bandwidth $\bigO(b)$, it follows that $\S(\lambda)\K(\btheta)^{-1}\S(\lambda)$, $\S(\lambda)^2$, $\R(\btheta)^2$, and  $\R(\btheta)\S(\lambda)$ in \eqref{eqn.llksum} are all banded matrices with bandwidth $\bigO(b)$. Thus, the computation of each quadratic form in the summand of \eqref{eqn.llksum} involves sparse matrix-vector multiplications and requires $\bigO(nnz)=\bigO(bN)$ operations. As a result, the computation of \eqref{eqn.llksum} has complexity $\bigO(bNT+kNT)$.

The second and third terms of the quasi log likelihood function \eqref{eqn.llk} involve the evaluation of two log determinants, $\log \det(\K(\btheta))$  and $\log |\det(\S(\lambda))|$, which is in general numerically unstable and computationally infeasible when the sample size $NT$ is large. To overcome such challenges, we utilize the relationship between an LU decomposition and the determinant. Recall that $\K(\btheta)$, given by $\sum_{j=0}^{\infty} \A(\btheta)^{j} \A(\btheta)^{'j}$, is dense in general. Thus,  it is computationally challenging to compute its log determinant and invert the matrix, as these operations involve solving large linear systems and infinite sum of matrices.  Here, we overcome the difficulty by taking full advantage of the symmetric spatial weight matrix and noting the following identity:  $\S(\lambda)\K(\btheta)^{-1}\S(\lambda) = \S(\lambda)^2 - \R(\btheta)^2$. After some algebra, we have,
\begin{equation*}
	\log(\det(\K(\btheta)))= \log \det(\S(\lambda)^2) - \log\det(\S(\lambda)^2 - \R(\btheta)^2),
\end{equation*}
which converts the computationally intensive task into sparse matrix multiplication and calculation of the (log-)determinant of two positive definite matrices with bandwidth $\bigO(b)$. Furthermore, incomplete LU  (ILU) decomposition of banded matrix takes advantage of the sparsity pattern to speed up the LU factorization without compromising the accuracy \citep{Saad2003}. This reduces the computational cost from the standard $\bigO(N^{2.4})$ to $\bigO(b^2N)$ \citep[see, e.g., Section 2 of ][]{kilic} and ensures the numerical stability of the calculation of log determinant of banded matrices during the evaluation of quasi log likelihood. 

The first term of the quasi log likelihood function \eqref{eqn.llk} would require $\bigO(N)$ operations after profiling out $\sigma^2$ in \eqref{eqn.llk}. That is, by setting
\begin{equation*} 
	\frac{\partial \log L_{NT}(\bdelta)}{\partial \sigma^2} = -\frac{NT}{2\sigma^2} + \frac{1}{2\sigma^4}\H(\bbeta, \btheta),
\end{equation*}
to zero, we have $\hat{\sigma}^2 =  (NT)^{-1}\H(\bbeta, \btheta)$, where $\H(\bbeta, \btheta)=(\Y-\X\bbeta)'\bSigma(\btheta)^{-1}(\Y-\X\bbeta)$. 
Combining the results above, the overall computational complexity for evaluating the quasi log likelihood function \eqref{eqn.llk} is $\bigO(bNT+kNT+b^2N)$. 

To compute the gradient  of \eqref{eqn.llk}, the computational cost is on the order $\bigO(kbNT + b^2N)$, because the partial derivative of  \eqref{eqn.llk} with respect to $\bbeta$ has a closed form 
\begin{eqnarray*}
	&&\frac{\partial  \log L_{NT}(\bdelta)}{\partial \bbeta} =  \\
	&&2\Big[
	\X_1'\S(\lambda)\K(\btheta)^{-1}\S(\lambda)(\X_1\bbeta - \Y_1) + \sum_{t=2}^T\X_t'\S(\lambda)^2(\X_t\bbeta - \Y_t)+\sum_{t=1}^{T-1}\X_t'\R(\btheta)^2(\X_t\bbeta - \Y_t) \nonumber\\
	&& +\sum_{t=1}^{T-1}\left\{\X_t'\R(\btheta)\S(\lambda)(\Y_{t+1} -\X_{t+1}\bbeta) 
	+ \X_{t+1}'\R(\btheta)\S(\lambda)(\Y_t - \X_t\bbeta) \right\}\Big] 
\end{eqnarray*}
and requires $\bigO(kbNT)$ operations, due to the multiplication of sparse matrices. The computational complexity of calculating the partial derivative of \eqref{eqn.llk} with respect to $\btheta$ using the analytical form remains computationally expensive as it involves solving large linear system requiring $\bigO(N^{2.4})$ operations. 
Thus, we use finite difference approximations in the gradient calculation, which reduce the computational cost from $\bigO(N^{2.4}+kbNT)$ to $\bigO(b^2N + kbNT)$. 

With the results above combined, the estimation of $\widehat\bdelta$ through numerical constrained optimization would require $\bigO(kbNT + b^2N)$ operations. In other words, the computational complexity of our method is linear to the total sample size ($NT$) when $k$ and $b$ are fixed and hence, is computationally feasible for large datasets even on the order of millions.

Last but not least, we turn to the computational cost involved in evaluating the estimate of $\var(\widehat{\bdelta})$.   By a similar argument in the evaluation of the quasi log likelihood, computing $\hat\sigma^{2}(\X'\bSigma^{-1}(\widehat\btheta)\X)^{-1}$ requires only $\bigO(kbNT)$ operations and $\bigO(kNT + bN)$ memory usage, as opposed to $\bigO(N^2T)$ operations and an extra $\bigO(N^2T)$ memory usage with the standard computation. Thus, our procedure facilitates the statistical inference about $\bbeta$ with large sample size. However,  the computation of \eqref{eqn.asymcovdelta} is dominated by solving a large linear system \citep{Gilbert92sparsematrices} in the calculation of $\bSigma_{1,N}$ and $\bSigma_{2,N}$
, which requires at most $\bigO(N^{2.4}T^{2.4})$ computations using the Coppersmith-Winograd algorithm \citep{Coppersmith}. As such, for practical applications, it may be prudent to apply resampling to compute the standard errors of the spatio-temporal dependence parameter estimates in $\widehat\btheta$. 
For example, spatial subsampling may be applied to overlapping or non-overlapping spatial blocks and provide replications of $\widehat{\btheta}$ for estimating the asymptotic covariance matrix \citep[see, e.g.,][]{sherman96, nordman2004}.

\section{Simulation Study} 
\label{sec:sim}

\subsection{Simulation Setup}

We conduct simulation experiments to assess the finite-sample properties of our proposed methodology and evaluate its computational efficiency. For the design matrix $\X$, we let $k=2$ including the intercept and a covariate sampled from the standard Gaussian distribution $\mathcal{N}(0,1)$. Once generated, $\X$ is kept fixed. The true parameter vector $\bdelta_0$ is set at $(1,0.5,0.1, 0.05, 0.7,-0.03)'$. The random innovations $\V_t$ are sampled independently from the standard Gaussian distribution, $t=1,\ldots,T$. We also consider a two-dimensional spatial domain with the data taken at spatial coordinates $\{(1,1),\ldots,(1,n),\ldots,(n,n)\}$ and the spatial weight matrix $\W$ is under a first-order spatial neighborhood structure. To examine the effect of sample sizes, we consider $N=n^2\in\{10^2, 20^2, 50^2, 100^2, 200^2\}$ and $T\in\{5,10,20,50\}$. For each combination of $N$ and $T$, 1000 simulations are generated.

The core computation is executed on an application server with dual Intel Xeon Silver 4116 2.1GHz 12-core (24 thread) processors and 512GB of RAM, running \textsc{Matlab} R2020a.



\subsection{Simulation Results}

The QMLE $\widehat{\bdelta}$ of the model parameter vector is obtained from maximizing the quasi log likelihood \eqref{eqn.llk}. To evaluate the finite-sample properties of the parameter estimates, we compute the bias and mean squared error (MSE) by taking the sample average of the differences and the squared differences between the estimate $\widehat{\bdelta}$ and the true value $\bdelta_0$ over the 1000 simulations for different \\
ations of $N$ and $T$ (Table \ref{tab:biasmse}). Overall, both the bias and the MSE decrease gradually as $N$ or $T$ increases for each of the parameters in $\bdelta_0$.

Next, we compare various estimates of the standard errors (SE) of the regression coefficients $\widehat{\bbeta}=(\hat{\beta}_0, \hat{\beta}_1)'$, which are of primary interest. Table \ref{tab:SE} shows the sample standard deviation (SD) of the estimates among 1000 simulations, the asymptotic SD approximated by $\sigma^{-2}_0N^{-1}\X'\bSigma(\btheta_0)^{-1}\X$, and the plug-in SE developed in Corollary \ref{cor.beta} evaluated $\widehat{\bbeta}$. The sample SD can be viewed as the gold standard. Both the asymptotic SD and the plug-in SE are close to the sample SD for different combinations of $N$ and $T$, supporting the results of Corollary \ref{cor.beta}. 

We also evaluate the distributions of the estimated regression coefficients $\widehat{\bbeta}$. Note that both $\left(\sigma_0^{-2}\X'\bSigma(\btheta_0)^{-1}\X\right)^{1/2}(\widehat{\bbeta}-\bbeta_0)$ and $\left(\hat{\sigma}^{-2}\X'\bSigma(\widehat\btheta)^{-1}\X\right)^{1/2}(\widehat{\bbeta}-\bbeta_0)$ converge in distribution to the standard bivariate Gaussian distribution by Corollary \ref{cor.beta} and the  Slutsky's theorem. Table \ref{tab:coverage} reports the coverage probabilities of the confidence intervals for $\beta_0$ and $\beta_1$ under the nominal level of 95\% using the asymptotic SD and the plug-in SE. The confidence intervals for $\beta_0$ and $\beta_1$ achieve the nominal coverage well for different combinations of $N$ and $T$.  In addition, for different $\bdelta_0$ and $\W$, the results are similar and not shown here to save space. 

The last column of Tables \ref{tab:biasmse} and \ref{tab:SE} reports the average time (in second) required to obtain the QMLE $\widehat\bdelta$ and the various measures of the variation of $\widehat{\bbeta}$.  The computation is reasonably fast. For example, when $N$ is large (e.g., $200^2$), the parameter estimation takes less than one minute per simulation. Moreover, the computational time is empirically linear to the spatial dimension $N$ as the length $T$ of the time series is relatively small. It is worthwhile to point out that the memory usage remains low (e.g., around 2GB when $N=200^2$ and $T=50$) in the computation. 

Overall, the simulation experiments corroborate the theoretical properties of $\widehat{\bdelta}$ and the computational complexity shown in Sections \ref{sec:theory} and \ref{sec:compute} respectively.


\section{Data Example: Land Surface Temperature} \label{sec:realdata}

As described in Section~\ref{sec:intro}, we regress the response variable of LST on the predictor variables of time trend, ecoregion classes, and interactions between the time trend and ecoregions, as well as the environmental covariates of elevation and latitude over $T=19$ years and $N=155,900$ image pixels per year.
Thus, there are a total of $k=171$ regression coefficients.
To implement the proposed spatio-temporal regression method, we construct a binary spatial weight matrix, $\W = (w_{ii'})_{N\times N}$, such that $w_{ii'}=1$ if cell $i'$ is a first-order neighbor of cell $i$, 0 otherwise. We then apply the computational procedure described in Section \ref{sec:compute}.  

\begin{figure}
	\begin{subfigure}{.33\textwidth}
		\centering
		\includegraphics[width=1\linewidth]{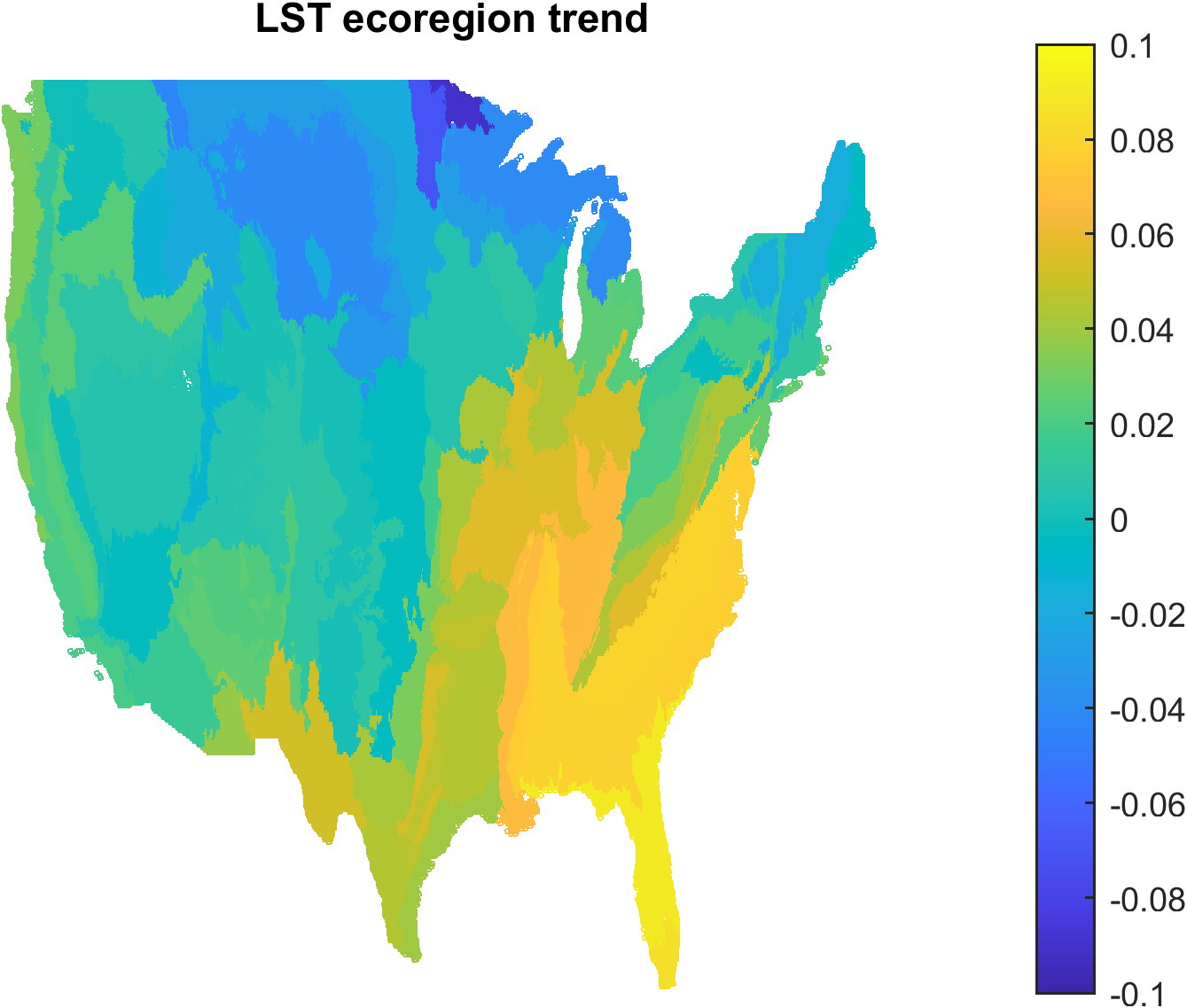}  
	\end{subfigure}
	\begin{subfigure}{.33\textwidth}
		\centering
		\includegraphics[width=1\linewidth]{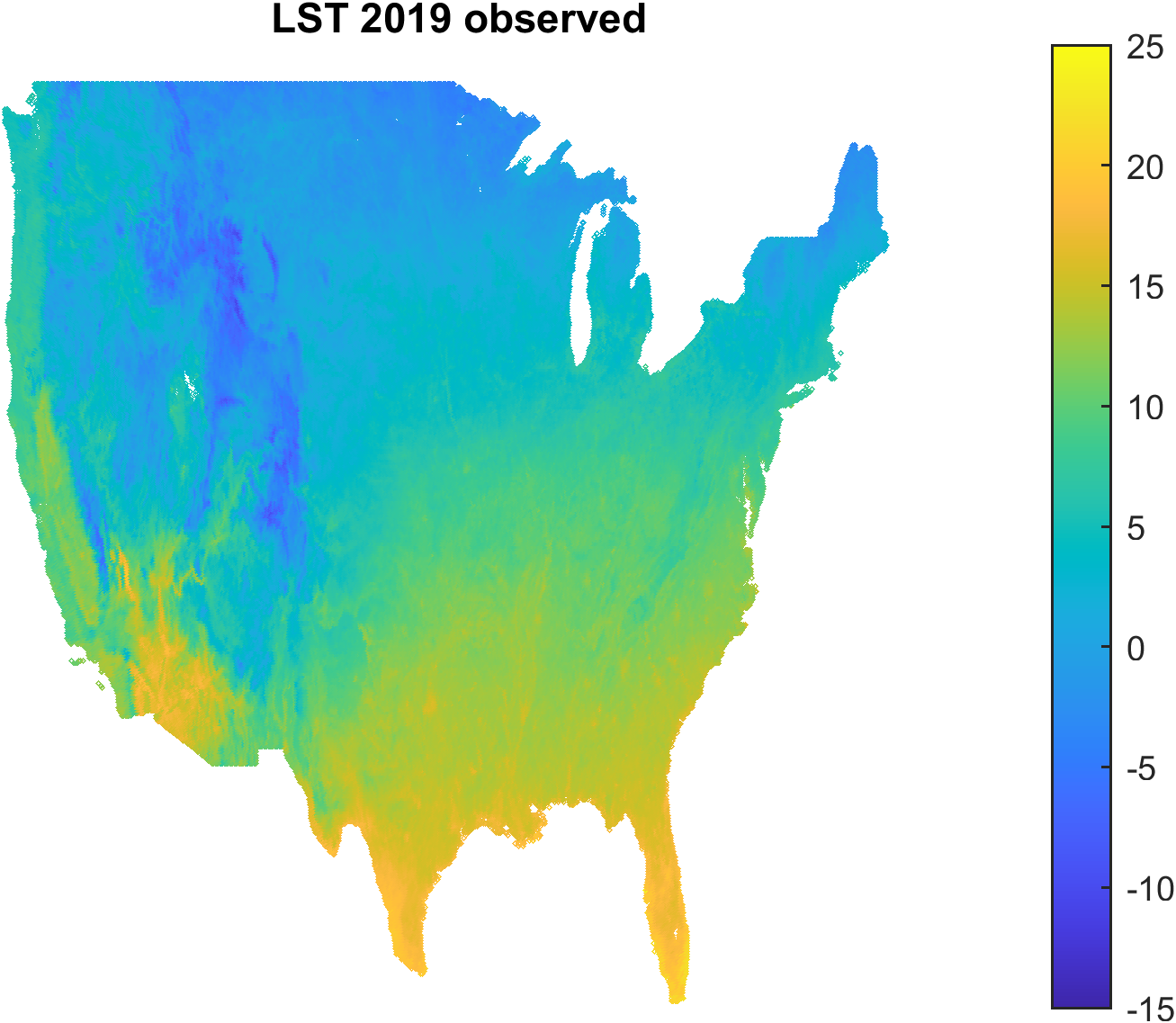}  
	\end{subfigure}
	\begin{subfigure}{.33\textwidth}
		\centering
		\includegraphics[width=1\linewidth]{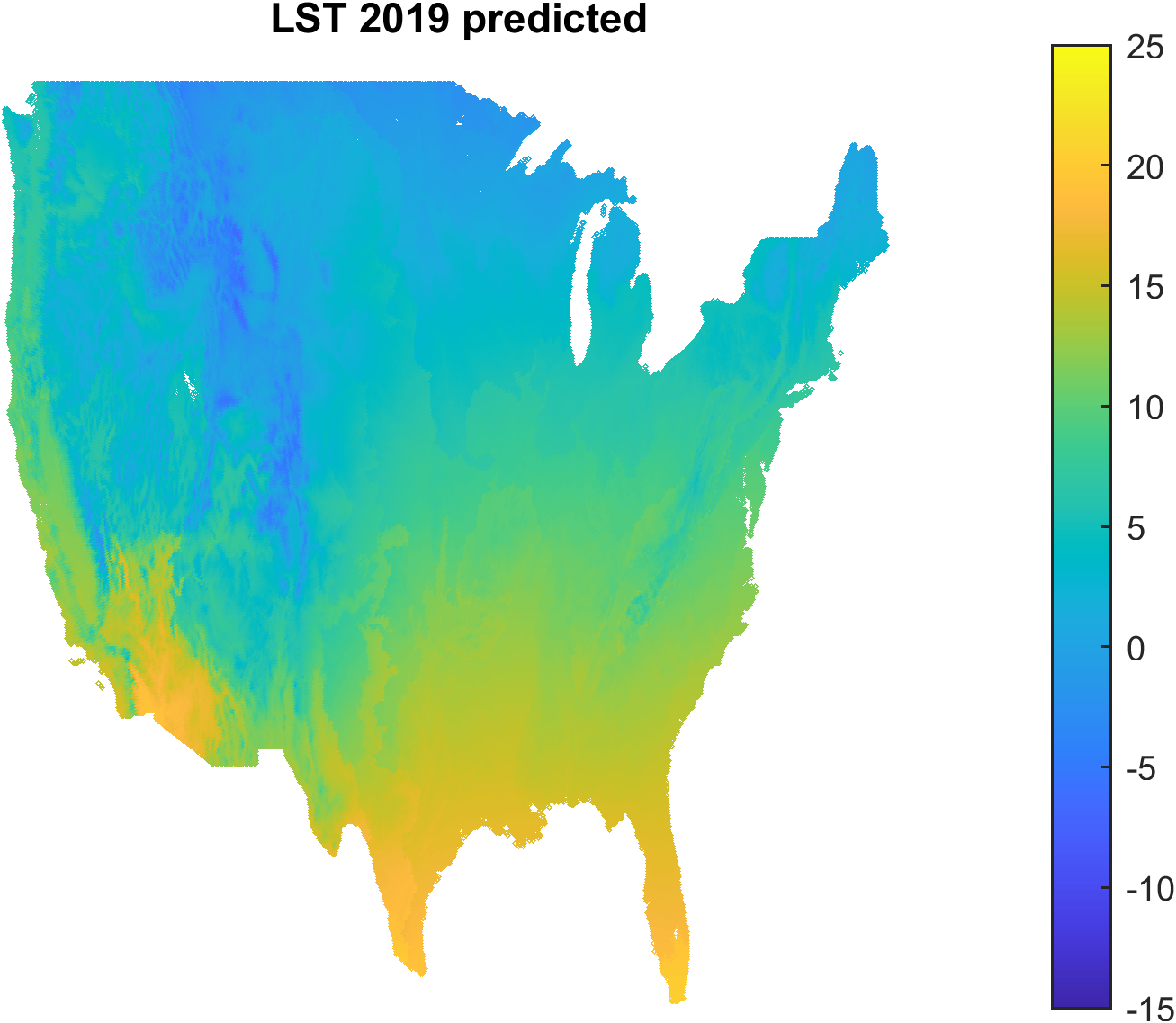}  
	\end{subfigure}
	\caption{Estimated regression coefficients by ecoregion using data from 2001 to 2019  (left panel); Observed land surface temperature (LST) in 2019 (middle panel); and Predicted LST in 2019 based on model fitting with data from 2001 to 2018 (right panel).}
	\label{fig:coef_out_sample}
\end{figure}

The majority of the regression coefficients are significant after false discovery rate adjustments, suggesting that, as expected, the mean LST values are different among different ecoregion classes and the time trend in LST varies among ecoregions (Figure \ref{fig:divergent_plot}). 
Left panel of Figure \ref{fig:coef_out_sample} maps the estimated time trend across ecoregions for the LST. 
Overall, there is an increasing time trend, especially in the southern and southeastern parts of the US, suggesting that these regions are subject to higher air temperatures than the rest of the continental US. 
This finding is consistent with previous findings that South and Southeast US seem to warm up the most in recent decades \citep{CSSR}. 
Tables \ref{table:LST_large} and \ref{table:LST_small} give the estimated regression coefficients of elevation, latitude, and the intercept (with respect to water), as well as the time trend of the five largest and smallest ecoregions respectively. 
The LST tends to decrease with elevation and latitude, which are as expected. The estimates for $\sigma^2, \lambda, \gamma$, and $\rho$ are $0.6061, 0.0360, 0.7273$, and $-0.0247$, respectively. 

For model diagnostics, we first assess the in-sample model fit. From Figure \ref{fig:Y_diff}, the estimated LST in 2001, 2019 and their difference  are similar to the observed data, indicating that our method can recover the mean function well using the covariates.
We then evaluate the out-of-sample prediction by fitting the 2001-2018 data and predicting the LST in 2019. The middle and right panels of Figure \ref{fig:coef_out_sample} suggest that the predicted LST for 2019 match up with the actual observations.


Finally, we compare our method with GpGp with the same set of covariates \citep{Guinness2021}.  The default neighborhood structure and the exponential space-time covariance function are adopted for fitting models using the R package {\tt GpGp}. 
The estimated LST values from GpGp seem to be quite different from the observed values (right panels of Figure \ref{fig:Y_diff}), possibly due to numerical instability with the large sample size $NT$.
While the computational complexity and the programming languages are not directly comparable between GpGp and our method, it took GpGp more than four days and our method within two hours to perform the regression analysis. 



\section*{Acknowledgments}
This research is supported by a NASA-AIST grant 80NSSC20K0282.


\FloatBarrier

\bibliographystyle{agsm}

\bibliography{reference_NASA}

%
\begin{appendices}
	
\section{Notation and Assumptions} 	\label{sec:assume}

We first introduce some notations and conventions. Given an $n\times n$ matrix $\P = (p_{ij})_{n\times n}$, we use $\tr(\P)$ and $\det(\P)$ to denote the trace and determinant of a square matrix $\P$, and we let $vec_D(\P)$ denote the column vector formed by the diagonal elements of $\P$. The $(i,j)$th element of a matrix $\P$ is denoted by $\ent_{ij}(\P)$. We define $\|\P\|_{1} = \max_{1\le j\le n}\sum_{i=1}^n|p_{ij}|$ and $\|\P\|_{\infty} = \max_{1\le i\le n}\sum_{j=1}^n|p_{ij}|$. We also let  $\|\P\|_2=\{\lambda_{\max}(\P'\P)\}^{1/2}$ and $\|\P\|_F=\{\tr(\P'\P)\}^{1/2}$ denote the spectral norm and the Frobenius norm, respectively. Let $abs(\P) = (|p_{i,j}|)_{n\times n}$. A sequence of $n\times n$ matrix $\P_n$ is said to be uniformly bounded (UB) in row and column sums, if $\sup_{n\ge1}\|\P_n\|_{1} <\infty$ and $\sup_{n\ge1}\|\P_n\|_{\infty} <\infty$. We also use $\zero$ and $\one$ to denote a matrix or a vector with all elements equal zero and one respectively.

For a real-valued function $f(\x)$, $\x = (\X_1, \ldots, x_k)'\in \mathbb{R}^k$, we let $\nabla f(\x)$ denote the gradient vector and let $\nabla^2 f(\x)$ denote the Hessian matrix. The partial derivative of $f$ with respect to $x_j$ is denoted by $\partial_{x_j} f(\x)$ or $\frac{\partial f(\x)}{\partial x_j}$, whereas the second partial derivative with respect to $x_j$ is denoted as $\partial_{x_jx_j} f(\x)$ (or $\frac{\partial^2 f(\x)}{\partial x_j^2}$).

In the following, we provide the regularity conditions for the establishing the large-sample properties of the QMLE $\widehat{\bdelta}$.

\begin{assume}\label{assume.W} 
	The $N\times N$ spatial weight matrix $\W$ is non-stochastic, symmetric, and the diagonal elements are zeros. 
\end{assume}

\begin{assume}\label{assume.S}
	The parameter space $\mathbf{\Theta}_{\bdelta}$ of $\bdelta= (\bbeta', \btheta', \sigma^2)'$ is compact and is the product space of $\mathbf{\Theta}_{\bbeta}$, $\mathbf{\Theta}_{\btheta}$ and $[\underline{\sigma}^2, \bar{\sigma}^2]$, where $\mathbf{\Theta}_{\btheta}$ is a compact set such that the matrices $\I_N -\lambda \W$ are nonsingular and the eigenvalues of $A(\btheta)$ are less than 1 in magnitude, while $\mathbf{\Theta}_{\bbeta}$ is a compact subset of $\mathbb{R}^k$. The true value $\bdelta_0= (\bbeta_0', \btheta_0', \sigma_0^2)'$ lies in the interior of $\mathbf{\Theta}_{\bdelta}$.
\end{assume}

\begin{assume}\label{assume.V} 
	The vector of innovations $ \V_t = (v_{1,t}, \ldots, v_{N,t})'$  $\sim iid(0, \sigma_0^2 \I_{N})$ and $E(|v_{j,t}|^{4+\eta}) <\infty$ for some $\eta>0$ for all $j,t$. 
\end{assume}

\begin{assume}\label{assume.UB} 
	The precision matrix, infinite sum of power of $\A(\btheta_0)$, and the design matrix 
	are uniformly bounded (UB).
	\begin{enumerate}[label=(\roman*)]
		\item $\bSigma(\btheta)^{-1}=\B (\btheta)'(\bOmega(\btheta))^{-1}\B (\btheta)$ and $\S(\lambda)^{-1}$ are UB, $\forall \btheta \in \mathbf{\Theta}$.
		\item $\sum_{h=1}^{\infty}abs(\A(\btheta_0)^h)$ is UB. 
		\item The $N\times k$ design matrix $\X_{t}$ is nonstochastic with elements UB in $N$ and $t$. 
	\end{enumerate}
\end{assume}

\begin{assume}\label{assume.x} 
	$\lim_{N\rightarrow\infty} \frac{1}{N}  \X'\bSigma(\btheta)^{-1}\X=\lim_{N\rightarrow\infty} \frac{1}{N}  \X'\B (\btheta)'(\bOmega(\btheta))^{-1}\B (\btheta)\X$ is nonsingular $\forall\btheta\in\mathbf{\Theta}$.  
\end{assume}


\begin{assume}\label{assume.sigma} 
	Denote by $\lambda_j(\btheta)$, $j=1, \ldots, NT$, the distinct eigenvalues of $\bSigma(\btheta)^{-1}\bSigma(\btheta_0)$ in non-increasing order. Let $f_j(\balpha) = -\log(\lambda_j(\btheta)\frac{\sigma_0^2}{\sigma^2}) + \lambda_j(\btheta)\frac{\sigma_0^2}{\sigma^2}$ where $\balpha = (\btheta', \sigma^2)'$, then $$\liminf_{N \rightarrow\infty} \frac{1}{N} \sum_{j=1}^{NT} \nabla^2 f_j(\balpha)$$ is nonsingular.
\end{assume}

\begin{assume}\label{assume.UB2} \
	$\bSigma(\btheta)$, $\partial_{\theta_i} (\bSigma(\btheta)^{-1})$, $\partial_{\theta_i\theta_j}^2 (\bSigma(\btheta)^{-1})$, and $\partial_{\theta_i\theta_j\theta_k}^3 (\bSigma(\btheta)^{-1})$  are UB in $\btheta = (\theta_1, \theta_2, \theta_3)' \in \mathbf{\Theta}$.
\end{assume}

\begin{assume}\label{assume.hessian} 
$\lim_{N\rightarrow\infty}N^{-1}\bOmega_{N} $ is nonsingular, where
\begin{align} 
	\bOmega_{N}   
	=&  \left(
	\begin{array}{cccc}
		\tr( \m_{\lambda}^2 )   &  \tr( \m_{\lambda} \m_{\gamma} ) &  \tr( \m_{\lambda} \m_{\rho} ) &  -\frac{1}{\sigma_0^2}  \tr(\m_{\lambda}) \\
		* &   \tr( \m_{\gamma}^2 )  & \tr( \m_{\gamma} \m_{\rho} ) &  -\frac{1}{\sigma_0^2}  \tr(\m_{\gamma})\\
		* &  * &    \tr( \m_{\rho}^2 )   & -\frac{1}{\sigma_0^2}  \tr(\m_{\rho}) \\
		* &  * &  * &  \frac{NT}{ \sigma_0^4 }  
	\end{array}
	\right) ,\label{eqn.bSigma1N}
\end{align} 
with $\m_{\lambda}$, $\m_{\gamma}$, and $\m_{\rho}$ defined in \eqref{eqn.Mlambda} in the Supplementary Materials.  
\end{assume}

\end{appendices}


\begin{landscape}
\begin{table}[ht]
	\centering
	\small
	\caption{Sample average bias ($\times 10^{-4}$) and mean squared error (MSE, $\times 10^{-4}$) of $\widehat{\bdelta}$ based on 1000 simulations, and average computational time (in second) per simulation.}
	\begin{tabular}{rr|rrrrrr|rrrrrr|r}
		\hline\hline
		&     & \multicolumn{6}{c|}{Average bias   $\times 10^{-4}$}         & \multicolumn{6}{c|}{Sample MSE $\times   10^{-4}$}           &  \begin{tabular}[c]{@{}c@{}}Average \\ time\end{tabular} \\
		$N$   & $T$   & $\beta_0$  & $\beta_1$  & $\lambda$ & $\gamma$   & $\rho$    & $\sigma^2$ & $\beta_0$  & $\beta_1$  & $\lambda$ & $\gamma$   & $\rho$    & $\sigma^2$  &  \\ \hline\hline
		$10^2$  & 5  & -22.80 & 9.21   & -5.24  & -50.05 & -20.98 & -105.50 & 68.32 & 16.42 & 2.60 & 17.73 & 5.20 & 40.92 & 0.05   \\
		& 10 & -14.23 & -10.61 & -11.03 & -35.19 & -6.29  & -56.64  & 34.28 & 8.32  & 1.51 & 8.42  & 2.51 & 20.72 & 0.05   \\
		& 20 & -17.43 & 12.26  & 1.40   & -15.47 & -7.07  & -45.87  & 18.00 & 4.35  & 0.66 & 3.90  & 1.38 & 9.99  & 0.07   \\
		& 50 & -9.01  & -1.41  & -2.64  & 1.08   & -1.06  & -10.59  & 7.99  & 1.51  & 0.27 & 1.46  & 0.48 & 3.99  & 0.11   \\ \hline
		$20^2$  & 5  & -23.22 & 14.89  & 2.23   & 3.27   & -12.63 & -45.95  & 15.87 & 4.49  & 0.63 & 4.00  & 1.38 & 10.66 & 0.10   \\
		& 10 & 16.95  & 0.09   & -0.69  & -10.73 & -1.35  & -6.48   & 8.68  & 2.09  & 0.32 & 1.90  & 0.67 & 5.25  & 0.12   \\
		& 20 & -0.21  & 1.18   & 1.11   & -9.17  & -1.73  & -8.68   & 4.99  & 1.01  & 0.16 & 0.93  & 0.29 & 2.58  & 0.17   \\
		& 50 & 3.58   & 1.33   & -0.19  & -1.55  & 0.24   & -2.55   & 1.89  & 0.41  & 0.07 & 0.40  & 0.10 & 0.94  & 0.24   \\ \hline
		$50^2$  & 5  & -0.55  & 2.01   & -0.32  & 0.82   & 0.00   & -4.31   & 2.57  & 0.65  & 0.11 & 0.61  & 0.22 & 1.69  & 0.43   \\
		& 10 & -0.85  & -0.58  & 0.94   & -0.26  & -2.54  & -9.74   & 1.34  & 0.34  & 0.05 & 0.31  & 0.09 & 0.83  & 0.54   \\
		& 20 & -3.11  & 0.41   & 1.16   & 0.14   & -1.53  & -0.69   & 0.75  & 0.16  & 0.02 & 0.15  & 0.05 & 0.40  & 0.65   \\
		& 50 & -1.87  & -0.64  & 2.06   & -0.45  & -1.31  & -1.51   & 0.29  & 0.06  & 0.01 & 0.06  & 0.02 & 0.16  & 0.96   \\ \hline
		$100^2$ & 5  & 5.11   & 1.19   & 1.16   & -1.89  & -0.57  & -0.08   & 0.65  & 0.16  & 0.03 & 0.17  & 0.05 & 0.42  & 1.77   \\
		& 10 & -0.85  & 0.27   & 1.86   & 0.73   & -0.65  & -1.17   & 0.34  & 0.08  & 0.01 & 0.08  & 0.02 & 0.19  & 2.12   \\
		& 20 & 1.58   & 1.03   & 1.52   & 0.21   & -0.98  & -0.60   & 0.19  & 0.04  & 0.01 & 0.04  & 0.01 & 0.10  & 2.67  \\
		& 50 & 1.81   & 0.43   & 1.78   & 0.59   & -1.20  & -1.27   & 0.08  & 0.02  & 0.00 & 0.01  & 0.00 & 0.04  & 4.41  \\ \hline
		$200^2$ & 5  & 3.33   & -0.71  & 1.52   & -0.53  & -1.50  & -0.42   & 0.17  & 0.04  & 0.01 & 0.04  & 0.01 & 0.11  & 10.13  \\
		& 10 & -0.50  & 0.63   & 1.58   & 0.05   & -1.35  & -1.80   & 0.09  & 0.02  & 0.00 & 0.02  & 0.01 & 0.06  & 11.67  \\
		& 20 & 0.13   & -0.37  & 1.75   & 0.15   & -0.89  & -1.61   & 0.05  & 0.01  & 0.00 & 0.01  & 0.00 & 0.02  & 13.84  \\
		& 50 & -1.18  & 0.02   & 1.84   & 0.13   & -0.88  & -0.82   & 0.02  & 0.00  & 0.00 & 0.00  & 0.00 & 0.01  & 20.29  \\ \hline\hline                                                     
	\end{tabular}
	\label{tab:biasmse}
\end{table}

\end{landscape}

\begin{table}[ht]
	\centering
	\caption{Sample standard deviation (Sample SD), asymptotic standard deviation (Asy SD) at $\bbeta_0$, and average standard error (Plug-in SE) by Corollary \ref{cor.beta} at $\widehat{\bbeta}$, based on 1000 simulations, and average computational time (in second) per simulation for Plug-in SE.}
	\small 
	\begin{tabular}{rr|rrrrrrr}
		\hline\hline
		$N$   & $T$   & \multicolumn{2}{c}{\begin{tabular}[c]{@{}c@{}}
				Sample SD  \\ $\times 10^{-2}$
		\end{tabular}} & \multicolumn{2}{c}{\begin{tabular}[c]{@{}c@{}}
				Asy SD\\ $\times 10^{-2}$
		\end{tabular}} & \multicolumn{2}{c}{\begin{tabular}[c]{@{}c@{}}
				Plug-in SE  \\ $\times 10^{-2}$
		\end{tabular}} & \begin{tabular}[c]{@{}c@{}}Average \\ time\end{tabular} \\ 
		&     & $\beta_0$              & $\beta_1$             & $\beta_0$              & $\beta_1$ & $\beta_0$              & $\beta_1$ &\\ \hline\hline
		$10^2$  & 5  & 8.266 & 4.053 & 7.979 & 4.067 & 7.907 & 4.044 & 0.002 \\
		& 10 & 5.856 & 2.884 & 5.954 & 2.936 & 5.897 & 2.930 & 0.003 \\
		& 20 & 4.241 & 2.083 & 4.331 & 2.097 & 4.323 & 2.094 & 0.004 \\
		& 50 & 2.826 & 1.228 & 2.793 & 1.244 & 2.792 & 1.243 & 0.008 \\ \hline
		$20^2$  & 5  & 3.978 & 2.116 & 4.001 & 2.085 & 3.992 & 2.080 & 0.005 \\
		& 10 & 2.943 & 1.447 & 2.984 & 1.424 & 2.980 & 1.423 & 0.006 \\
		& 20 & 2.235 & 1.005 & 2.170 & 0.999 & 2.168 & 0.999 & 0.011 \\
		& 50 & 1.373 & 0.638 & 1.397 & 0.634 & 1.397 & 0.634 & 0.021 \\ \hline
		$50^2$  & 5  & 1.603 & 0.805 & 1.604 & 0.825 & 1.605 & 0.825 & 0.023 \\
		& 10 & 1.157 & 0.581 & 1.194 & 0.570 & 1.193 & 0.570 & 0.031 \\
		& 20 & 0.866 & 0.394 & 0.868 & 0.401 & 0.868 & 0.401 & 0.047 \\
		& 50 & 0.538 & 0.246 & 0.559 & 0.252 & 0.559 & 0.252 & 0.083 \\ \hline
		$100^2$ & 5  & 0.807 & 0.397 & 0.803 & 0.414 & 0.803 & 0.414 & 0.088 \\
		& 10 & 0.584 & 0.281 & 0.597 & 0.286 & 0.598 & 0.286 & 0.118 \\
		& 20 & 0.436 & 0.194 & 0.434 & 0.201 & 0.435 & 0.201 & 0.186 \\
		& 50 & 0.277 & 0.126 & 0.279 & 0.126 & 0.280 & 0.126 & 0.373 \\ \hline
		$200^2$ & 5  & 0.406 & 0.208 & 0.401 & 0.207 & 0.402 & 0.207 & 0.520 \\
		& 10 & 0.301 & 0.143 & 0.299 & 0.143 & 0.299 & 0.143 & 0.636 \\
		& 20 & 0.218 & 0.099 & 0.217 & 0.100 & 0.217 & 0.100 & 0.778 \\
		& 50 & 0.140 & 0.063 & 0.140 & 0.063 & 0.140 & 0.063 & 1.465  \\ \hline \hline 
	\end{tabular}
	
	\label{tab:SE}
\end{table}

\begin{table}[ht]
	\centering
	\caption{Coverage probabilities of the confidence intervals for $\beta_0$ and $\beta_1$ under the nominal level of 95\% using the asymptotic standard deviation (Asy SD) at $\bbeta_0$ and the average standard error (Plug-in SE) by Corollary \ref{cor.beta} at $\widehat{\bbeta}$ based on 1000 simulations.  }
	\centering
	\begin{tabular}{rr|rrrr}
		\hline\hline
		$N$   & $T$   & \multicolumn{2}{c}{Asy SD} & \multicolumn{2}{c}{Plug-in SE} \\
		&     & $\beta_0$        & $\beta_1$        & $\beta_0$          & $\beta_1$         \\ \hline\hline
		$10^2$  & 5  & 94.2 & 94.9 & 93.1 & 94.7 \\
		& 10 & 95.6 & 95.6 & 95.3 & 95.3 \\
		& 20 & 95.3 & 95.2 & 95.4 & 95.1 \\
		& 50 & 95.3 & 94.5 & 95.3 & 94.5 \\ \hline
		$20^2$  & 5  & 95.8 & 94.1 & 95.1 & 94.0 \\
		& 10 & 95.9 & 93.7 & 96.2 & 93.7 \\
		& 20 & 94.4 & 95.0 & 94.2 & 95.1 \\
		& 50 & 94.7 & 95.1 & 94.8 & 95.1 \\ \hline
		$50^2$  & 5  & 94.0 & 95.8 & 94.0 & 95.8 \\
		& 10 & 96.1 & 94.5 & 96.1 & 94.6 \\
		& 20 & 94.7 & 96.2 & 94.5 & 96.2 \\
		& 50 & 95.5 & 96.2 & 95.5 & 96.2 \\ \hline
		$100^2$ & 5  & 94.6 & 96.5 & 94.5 & 96.5 \\
		& 10 & 95.3 & 95.4 & 95.4 & 95.4 \\
		& 20 & 95.9 & 96.1 & 95.9 & 96.1 \\
		& 50 & 94.9 & 95.1 & 95.0 & 95.1 \\ \hline
		$200^2$ & 5  & 94.6 & 95.0 & 94.6 & 94.9 \\ 
		& 10 & 94.5 & 96.0 & 94.6 & 96.0 \\
		& 20 & 94.9 & 94.9 & 94.9 & 94.9 \\
		& 50 & 94.9 & 94.6 & 94.9 & 94.6     \\ \hline\hline   
	\end{tabular}	
	\label{tab:coverage}
\end{table}

\FloatBarrier

\end{document}